\documentclass[a4paper,UKenglish]{lipics-v2016}

\usepackage{microtype}
\usepackage[utf8]{inputenc}

\newcommand{\bb}{{\it BBhash}\xspace}

\title{Fast and scalable minimal perfect hashing for massive key sets}

\author[1]{Antoine Limasset}
\author[1]{Guillaume Rizk}
\author[2]{Rayan Chikhi}
\author[1]{Pierre Peterlongo}
\affil[1]{IRISA Inria Rennes Bretagne Atlantique, GenScale team, Campus de Beaulieu 35042 Rennes, France}
\affil[2]{CNRS, CRIStAL, Université de Lille, Inria Lille - Nord Europe, France}

\authorrunning{A. Limasset, G. Rizk, R. Chikhi, P. Peterlongo} 

\Copyright{Antoine Limasset, Guillaume Rizk, Rayan Chikhi and Pierre Peterlongo}

\usepackage{ amssymb }
\usepackage{textcomp}
\usepackage{graphicx}
\usepackage{amsmath}
\usepackage{url}
\usepackage[ backgroundcolor=cyan,linecolor=black]{todonotes}
\usepackage[noend,ruled,vlined]{algorithm2e}
\usepackage{tikz}
\usepackage{comment}
\usepackage{amsfonts}

\newcommand{\hf}{\texttt{MPHF}\xspace}

\newcommand\rc[1]{\todo[author=rayan, fancyline]{#1}}

\newtheorem{lem}{Lemma}
\makeatletter
\newcommand{\algorithmfootnote}[2][\footnotesize]{%
  \let\old@algocf@finish\@algocf@finish
  \def\@algocf@finish{\old@algocf@finish
    \leavevmode\rlap{\begin{minipage}{\linewidth}
    #1#2
    \end{minipage}}%
  }%
}
\makeatother
\begin{document}

\maketitle


\begin{abstract}
Minimal perfect hash functions provide space-efficient and collision-free hashing on static sets. Existing algorithms and implementations that build such functions have practical limitations on the number of input elements they can process, due to high construction time, RAM or external memory usage.
We revisit a simple algorithm and show that it is highly competitive with the state of the art, especially in terms of construction time and memory usage. We provide a parallel C++ implementation called \bb. It is capable of creating a minimal perfect hash function of $10^{10}$ elements in less than 7 minutes using 8 threads and 5 GB of memory, and the resulting function uses 3.7 bits/element. To the best of our knowledge, this is also the first implementation that has been successfully tested on an input of cardinality $10^{12}$.
Source code: \url{https://github.com/rizkg/BBHash}
\end{abstract}


\section{Introduction}
Given a set $S$ of $N$ elements (\emph{keys}), a minimal perfect hash function (\hf) is an injective function that maps each key of $S$ to an integer in the interval $[1,N]$.
In other words, an \hf labels each key of $S$ with integers in a collision-free manner, using the smallest possible integer range. A remarkable property is the small space in which these functions can be stored: only a couple of bits per key, independently of the size of the keys. Furthermore, an \hf query is done in constant time. While an \hf could be easily obtained using a key-value store (e.g. a hash table), such a representation would occupy an unreasonable amount of space, with both the keys and the integer labels stored explicitly.

The theoretical minimum  amount  of  space  needed  to  represent  an \hf is  known  to  be $\log_2(e)N\approx1,44N$ bits~\cite{fredman1984size,mehlhorn1982program}.  
In practice, for large key sets (billions of keys), many implementations achieve less than $3N$ bits per key, independently of the number of keys~\cite{belazzougui2009hash,czech1997perfect}. However no implementation comes asymptotically close to the lower bound for large key sets.
Given that {\hf}s are typically used to index huge sets of strings, e.g. in bioinformatics~\cite{chapman2011meraculous,Chen2013,chikhi2016compacting}, in network applications~\cite{lu2006perfect}, or in databases~\cite{Chang01012005}, lowering the representation space is of interest.
We observe that in many of these applications, {\hf}s are actually used to construct static dictionaries, i.e. key-value stores where the set of keys is fixed and never updated~\cite{chapman2011meraculous,chikhi2016compacting}.
Assuming that the user only queries the \hf to get values corresponding to keys that are guaranteed to be in the static set, the keys themselves do not necessarily need to be stored in memory. However the associated values in the dictionary typically do need to be stored, and they often dwarf the size of the \hf. The representation of such dictionaries then consists of two components: a space-efficient \hf, and a relatively more space-expensive set of values. In such applications, whether the \hf occupies 1.44 bits or 3 bits per key is thus arguably not a critical aspect.

In practice, a significant bottleneck for large-scale applications is the construction step of {\hf}s, both in terms of memory usage and computation time. 
Constructing {\hf}s efficiently is an active area of research. Many recent \hf construction algorithms are  based  on  efficient peeling of hypergraphs~\cite{belazzougui2014cache,botelho2007simple,botelho2013practical,Genuzio2016}. However, they require an order of magnitude more space during construction than for the resulting data structure.
For billions of keys, while the \hf itself can easily fit in main memory of a commodity computer, 
its construction algorithm requires large-memory servers.
To address this, Botelho and colleagues~\cite{botelho2013practical} propose to divide the problem by building many smaller {\hf}s, while Belazzougui \textit{et al.}~\cite{belazzougui2014cache} propose an external-memory algorithm for hypergraph peeling. Very recently, Genuzio \textit{et al.}~\cite{Genuzio2016} demonstrated practical improvements to the Gaussian elimination technique, that make it competitive with~\cite{belazzougui2014cache} in terms of construction time, lookup time and space of the final structure.
These techniques are, to  the  best  of  our  knowledge, the  most  scalable solutions available. However, when evaluating existing implementations, the construction of {\hf}s for sets that significantly exceed a billion of keys remains prohibitive in terms of time and space usage.  

A simple idea has been explored by previous works~\cite{chapman2011meraculous,lu2006perfect,Muller2014} for constructing {\hf}s using arrays of bits, or fingerprints. However, it has received relatively less attention compared to other hypergraph-based methods, and no implementation is publicly available in a stand-alone \hf library. In this article we revisit this idea, and introduce novel contributions: a careful analysis of space usage during construction, and an efficient, parallel implementation along with an extensive evaluation with respect to the state of the art. We show that it is possible to construct an \hf using nearly as much memory as the space required by the final structure, without partitioning the input. 
We propose a novel implementation called \bb (``Basic Binary representAtion of Successive Hashing'') 
with the following features:
\begin{itemize}	
\item construction space overhead is small compared to the space occupied by the \hf, 
\item multi-threaded,
\item scales up to to very large key sets (tested with up to 1 trillion keys).
\end{itemize}

To the best of our knowledge, there does not exist another usable implementation that satisfies any two of the features above.
Furthermore, the algorithm enables a time/memory trade-off: faster construction and faster query times can be obtained at the expense of a few more bits per element in the final structure and during construction.
We created an \hf for ten billion keys in 6 minutes 47 seconds and less than 5 GB of working memory, and an \hf for a trillion keys in less than 36 hours and 637 GB memory. Overall, with respect to others available \hf construction approaches, our implementation is at least two orders of magnitudes more space-efficient when considering internal and external memory usage during construction, 
and at least one order of magnitude faster.
The resulting \hf has slightly higher space usage and faster or comparable query times than other methods.


\section{Efficient construction of minimal perfect hash function}

\subsection{Method overview}

\begin{figure}[t]
  \centering
  \includegraphics[width=\textwidth]{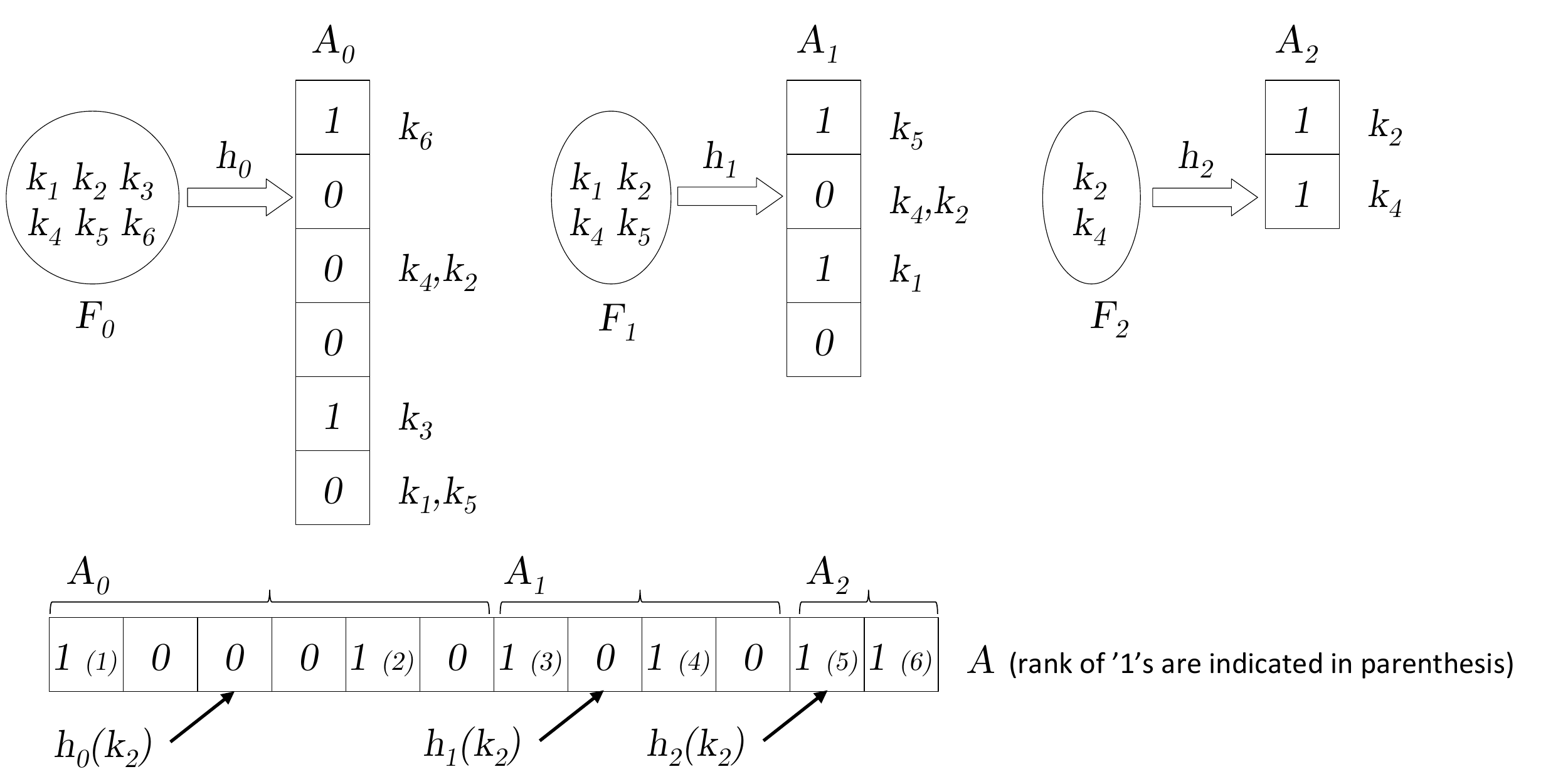}
  \caption{\hf construction and query example. The input is a set $F_0$ composed of $N=6$ keys ($k_1$ to $k_6$). All keys are hashed using a hash function $h_0$ and are attempted to be placed in an array $A_0$ at positions given by the hash function. The keys $k_3$ and $k_6$ do not have collisions in the array, thus the corresponding bits in $A_0$ are set to '1'. The other keys from $F_0$ that are involved in collisions are placed in a new set $F_1$. In the second level, keys from $F_1$ are hashed using a hash function $h_1$. Keys $k_1$ and $k_5$ are uniquely placed while $k_2$ and $k_4$ collide, thus they are then stored in the set $F_2$. With the hash function $h_2$, the keys from $F_2$ have no collision, and the process finishes. The \hf query operation is very similar to the construction algorithm. Let $A$  be the the concatenation of $A_0,A_1,A_2$ (see bottom part of the figure). To query $k_2$, the key is first hashed with $h_0$. The associated value in $A_0$ is '0', so $k_2$ is then hashed with $h_1$. The value associated in $A_1$ is again '0'. When finally hashed with $h_2$, the value associated in $A_2$ is '1' and thus the query stops here. The index returned by the \hf is the rank of this '1' (here, 5) in $A$. In this example, the \hf values returned when querying $k_1,k_2,k_3,k_4,k_5$ and $k_6$ are respectively 4,5,2,6,3, and 1.} 
  \label{fig:mphf_construction}
\end{figure}

Our \hf construction procedure revisits previously published techniques~\cite{chapman2011meraculous,lu2006perfect}. 
Given a set $F_0$ of keys, a classical hash function $h_0$ maps keys to an integer in $[1,|F_0|]$. 
A bit array $A_0$ of size $|F_0|$ is created such that there is a 1 at position $i$ if and only if exactly one element of $F_0$ has a hash value of $i$. We say that there is a \emph{collision} whenever two keys in $F_0$ have the same hash value.
Keys from $F_0$ that were involved in a collision are inserted into a new set $F_{1}$. 
The process repeats with $F_{1}$ and a new hash function $h_1$. 
A new bit array $A_1$ of size $|F_1|$ is created using the same procedure as for $A_0$ (except that $F_1$ is used instead of $F_0$, and $h_1$ instead of $h_0$). The process is repeated with $F_2,F_3,\ldots$ until one of these sets, $F_{\text{last}+1}$, is empty.

We obtain an \hf by concatenating the bit arrays $A_0, A_1, \dots, A_\text{last}$ into an array $A$. 
To perform a query, a key is hashed successively with hash functions $h_0,h_1,\ldots$ as long as the value in $A_i$ ($i\geq 0$) at the position given by the hash function $h_i$ is 0. Eventually, by construction, we reach a 1 at some position of $A$ for some $i=d$. We say that the \textit{level} of the key is $d$. The index returned by the \hf is the rank of this one in $A$. See Figure~\ref{fig:mphf_construction} for an example.

\subsection{Algorithm details}

\subsubsection{Collision detection} During construction at each level $d$, collisions are detected using a temporary bit array $C_d$ of size $|A_d|$. Initially all $C_d$ bits are set to '0'. A bit of $C_d[i]$ is set to '1' if two or more keys from $F_d$ have the same value $i$ given by hash function $h_d$. Finally, if $C_d[i]=1$, then $A_d[i]=0$. Formally:
\begin{align*}
 C_d[i]=1 &\Rightarrow A_d[i]=0;  \\
 (h_d[x]=i \text{ and } A_d[i]=0 \text{ and } C_d[i]=0) &\Rightarrow A_d[i]=1 \left( \text{and } C_d[i]=0\right);  \\
 (h_d[x]=i \text{ and } A_d[i]=1 \text{ and } C_d[i]=0) &\Rightarrow A_d[i]=0 \text{ and } C_d[i]=1.  
\end{align*}

\subsubsection{Queries}
A query of a key $x$ is performed by finding the smallest $d$ such that $A_d[h_{d}(x)]=1$. The (non minimal) hash value of $x$ is then $(\sum_{i<d} |F_i|) + h_{d}(x)$.

\subsubsection{Minimality}
To ensure that the image range of the function is $[1,|F_0|]$, we compute the cumulative rank of each '1' in the bit arrays $A_i$.
Suppose, that $d$ is the smallest value such that $A_d[h_{d}(x)]=1$. The minimal perfect hash value is given by $\sum_{i<d} (weight(A_i) + rank(A_d[h_{d}(x)])$, where ${weight}(A_i)$ is the number of bits set to '1' in the $A_i$ array, and $rank(A_d[y])$ is the number of bits set to 1 in $A_d$ within the interval $[0,y]$, thus $rank(A_d[y]) = \sum_{j<y}{A_d[j]}$. 
This is a classic method also used in other MPHFs~\cite{botelho2007simple}.

\subsubsection{Faster query and construction times (parameter $\gamma$)}
The running time of the construction depends on the number of collisions on the $A_d$ arrays, at each level $d$. One way to reduce the number of collisions, hence to place more keys at each level, is to use bit arrays ($A_d$ and $C_d$) larger than $|F_d|$. We introduce a parameter $\gamma \in \mathbb{R}$, $\gamma \geq 1$, such that $|C_d| = |A_d| = \gamma|F_d|$. With $\gamma=1$, the size of $A$ is minimal. With $\gamma\geq2$,  the number of collisions is significantly decreased and thus construction and query times are reduced, at the cost of a larger \hf structure size. The influence of $\gamma$ is discussed in more details in the following analyses and results.

\subsection{Analysis}
\label{sec:analysis}
Proofs of the following observations and lemma are given in the Appendix.  

\subsubsection{Size of the \hf}
The expected size of the structure can be determined using a simple argument, previously made in~\cite{chapman2011meraculous}. When $\gamma=1$, the expected number of keys which do not collide at level $d$ is $ |A_d| e^{-1}$, thus $|A_{d}| = |A_{d-1}| (1-e^{-1}) =  |A_0| (1-e^{-1})^{d}$. 
In total, the expected number of bits required by the hashing scheme is $\sum_{d\geq 0} |A_d| = N \sum_{d\geq 0} (1-e^{-1})^{d} = e N$, with $N$ being the total number of input keys ($N=|F_0|$).
Note that consequently the image of the hash function is also in $[1,eN]$, before minimization using the rank technique.
When $\gamma\geq1$, the expected proportion of keys without collisions each level $d$ is $|A_d| e^{-\frac{1}{\gamma}}$. Since each $A_d$ no longer uses one bit per key but $\gamma$ bits per key, the expected total number of bits required by the \hf is $\gamma e^{\frac{1}{\gamma}}N$.

\subsubsection{Space usage during construction}\label{sec:space-construction}
We analyze the disk space during construction. Recall that during construction of level $d$, a bit array $C_d$ of size $|A_d|$ is used to record collisions. Note that the $C_d$ array is only needed during the $d$-th level. It is deleted before level $d+1$. The total memory required during level $d$ is $\sum_{i\leq d}(|A_i|) + |C_d| = \sum_{i< d}(|A_i|) + 2|A_d|$.

\begin{lem}
For $\gamma>0$, the space of our \hf is $S= \gamma e^{\frac{1}{\gamma}}N$ bits. The maximal space during construction is $S$ when $\gamma \leq \log(2)^{-1}$, and $2S$ bits otherwise.
\end{lem}

A full proof of the Lemma is provided in the Appendix.



\section{Implementation}\label{sec:impl}
We present \bb, a C++ implementation 
available at \url{http://github.com/rizkg/BBHash}. 
We describe in this section some  design key choices and optimizations.

\subsection{Rank structure} 
We use a classical technique to implement the rank operation: the ranks of a fraction of the '1's present in $A$ are recorded, and the ranks in-between are computed dynamically using the recorded ranks as checkpoints.

In practice 64 bits integers are used for counters, which is enough for realistic use of an \hf, and placed every 512 positions by default.
These values were chosen as they offer a good speed/memory trade-off, increasing the size of the MPHF by a factor 1.125 while achieving good query performance. The total size of the MPHF is thus $(1+\frac{64}{512})\gamma e^{\frac{1}{\gamma}}N$

\subsection{Parallelization}\label{sec:para}

Parallelization is achieved by partitioning keys over several threads. The algorithm presented in Section 2 is executed on multiple threads concurrently, over the same memory space. 
Built-in compiler functions (e.g. \textit{sync\_fetch\_and\_or}) are used for concurrent access in the $A_i$ arrays. The efficiency of this parallelization scheme is shown in the Results section, but note that it is fundamentally limited by random memory accesses to the $A_i$ arrays which incur cache misses.

\subsection{Hash functions}
The \hf construction requires classical hash functions. Other authors have observed that common hash functions behave practically as well as fully random hash functions~\cite{belazzougui2009hash}.
We therefore choose to use xor-shift based hash functions~\cite{marsaglia2003xorshift} for their efficiency both in terms of computation speed and distribution uniformity~\cite{mitzenmacher2008simple}. 

\subsection{Disk usage}
 \label{ssec:diskusage}

In the applications we consider, key sets are typically too big to fit in RAM. Thus we propose to read them on the fly from disk. 
There are mainly two distinct strategies regarding the disk usage during construction: 1/ during each level $d$, keys that are to be inserted in the set $F_{d+1}$ are written directly to disk. The set $F_{d+1}$ is then read during level $d+1$ and erased before level $d+2$; or 2/ at each level all keys from the original input key file are read and queried in order to determine which keys were already assigned to a level $i<d$, and which would belong to $F_d$. 

The first strategy obviously provides faster construction at the cost of temporary disk usage. At each level $d>0$, two temporary key files are stored on disk: $F_d$ and $F_{d+1}$. The highest disk usage is thus achieved during level $1$, i.e. by storing $|F_1|+|F_2|=|F_0|((1-e^{-1/\gamma})+(1-e^{-1/\gamma})^2)$ elements. With $\gamma=1$, this represents $ \approx 1.03 N$ elements, thus the construction overhead on disk is approximately the size of the input key file. Note that with $\gamma=2$ (resp. $\gamma=5$), this overhead diminishes and becomes a ratio of $\approx 0.55$ (resp. $\approx 0.21$) the size of the input key file.

The first strategy is the default strategy proposed in our implementation. The second one has also been implemented and can be optionally switched on.




\subsection{Termination}
\label{ssec:termination}
The expected number of unplaced keys decreases exponentially with the number of levels but is not theoretically guaranteed to reach zero in a finite number of steps. To ensure termination of the construction algorithm, in our implementation a maximal number $D$ of levels is fixed. Then, the remaining keys are inserted into a regular hash table. Value $D$ is  a parameter, its default value is $D=25$ for which the expected number of keys stored in this hash table is $\approx 10^{-5} N$ for $\gamma=1$ and becomes in practice negligible for $\gamma\geq 2$, allowing the size overhead of the final hash table to be negligible regarding the final \hf size.

\section{Results}

We evaluated the performance \bb for the construction of large {\hf}s. We generated files containing various number of keys (from 1 million to 1 trillion keys). In our tests, a key is a binary representation of a pseudo-random positive integer in $[0;2^{64}]$. Within each file, each key is unique. We also performed a test where input keys are strings (n-grams) to ensure that using integers as keys does not bias results.
Tests were performed on a cluster node with a Xeon$^\text{\textcopyright}$ E5 2.8 GHz 24-core CPU, 256 GB of memory, and a mechanical hard drive. 
Except for the experiment with $10^{12}$ keys, running times include the time needed to read input keys from disk. Note that files containing key sets may be cached in memory by the operating system, and all evaluated methods benefit from this effect during \hf construction.
We refer to the Appendix for the specific commands and parameters used in these experiments.

We first analyzed the influence of the  $\gamma$ value (the main parameter of \bb), then the effect of using multiple threads depending on the parallelization strategy. Second, we compared \bb with other state-of-the-art methods. Finally, we performed an \hf construction on $10^{12}$ elements.

\subsection{Influence of the $\gamma$ parameter}
\label{ssec:gamma}
\begin{figure}
  \centering
  \includegraphics[width=0.49\textwidth]{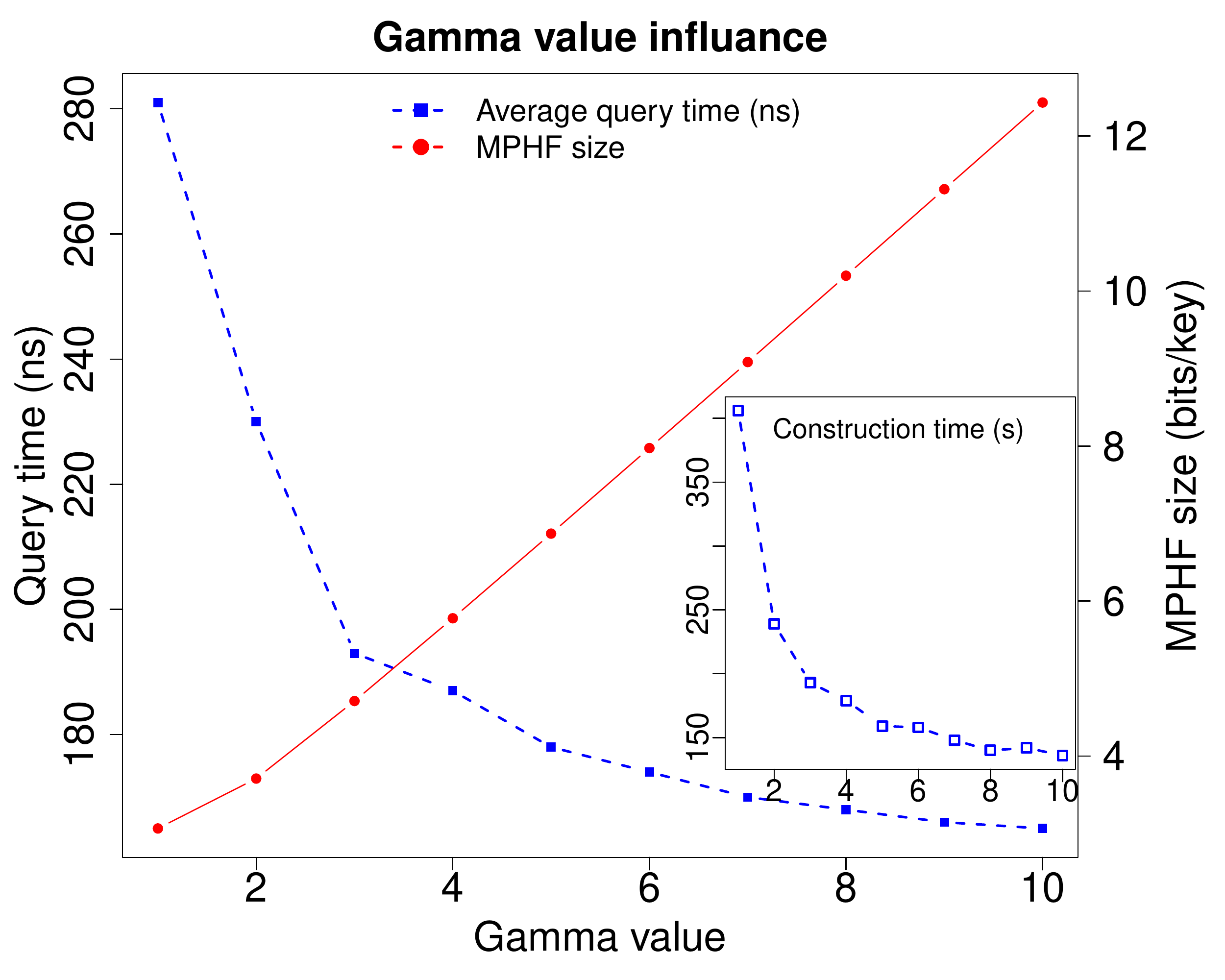}
  \includegraphics[width=0.49\textwidth]{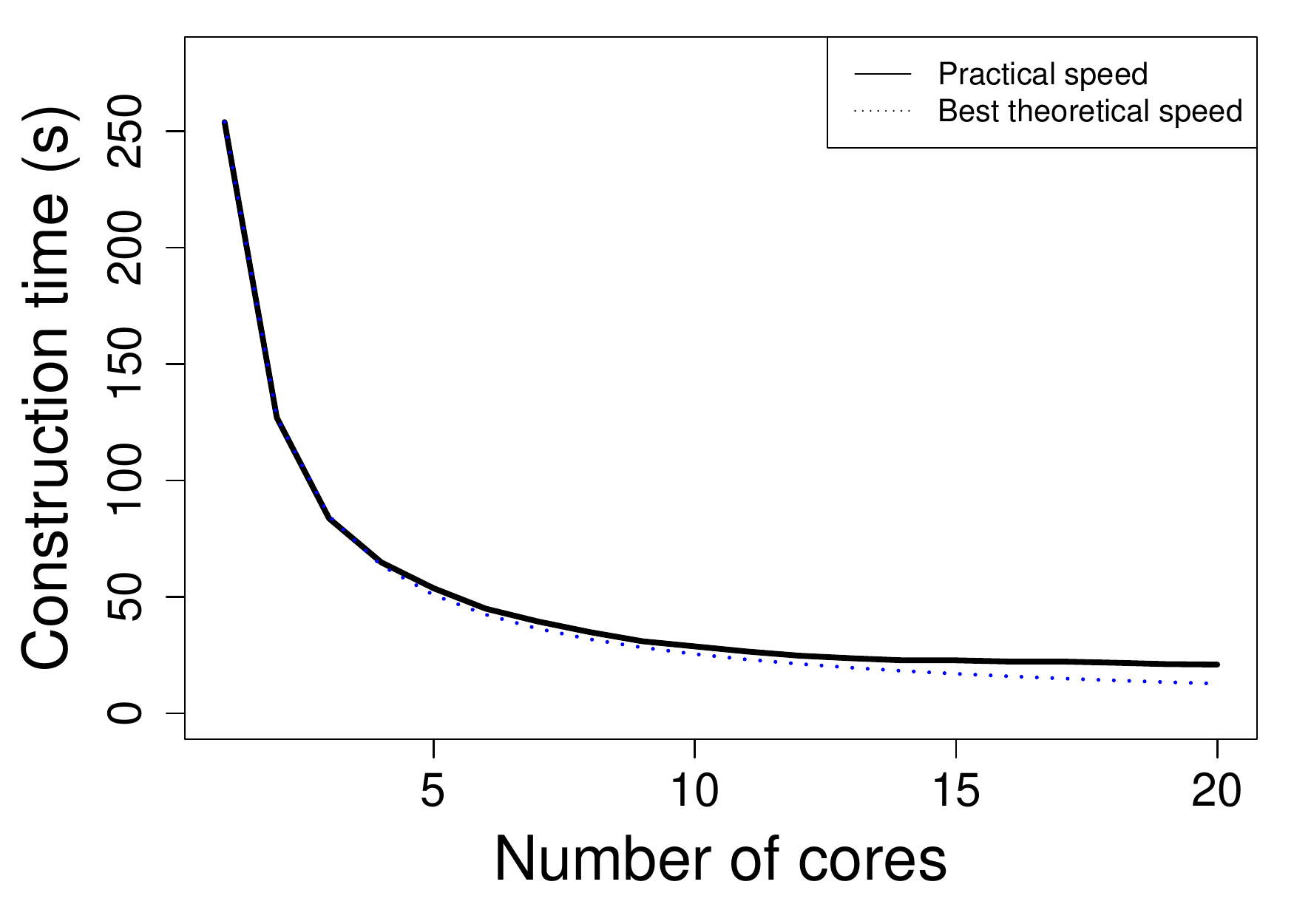}
  \caption{Left: Effects of the gamma parameter on the performance of \bb when run on a set composed of one billion keys, when executed on a single CPU thread.. Times and \hf size behave accordingly to the theoretical analysis, respectively  $O(e^{(1/\gamma)})$, and $O(\gamma e^{(1/\gamma)})$.
  Right: Performance of the \bb construction time according to the number of cores. Tests were performed on an input set composed of one billion keys, using $\gamma=2$. The ``\textit{best theoretical speed}'' curve is computed by dividing the construction time obtained with one core by the number of cores.}
   \label{fig:gamma}
\end{figure}

We report in Figure~\ref{fig:gamma} (left) the construction times and the mean query times, as well as the size of the produced \hf, with respect to several $\gamma$ values. 
The main observation is that $\gamma \geq 2$ drastically accelerates construction and query times.
This is expected since large $\gamma$ values allow more elements to be placed in the first levels of the \hf; thus limiting the number of times each key is hashed to compute its level. In particular, for keys placed in the very first level, the query time is limited to a single hashing and a memory access. The average level of all keys is $e^{(1/\gamma)}$, we therefore expect construction and query times to decrease when $\gamma$ increases.
However, larger $\gamma$ values also incur larger \hf sizes. One observes that $\gamma>5$ values seem to bring very little advantage at the price of higher space requirements. A related work used $\gamma=1$ in order to minimize the \hf size~\cite{chapman2011meraculous}. Here, we argue that using $\gamma$ values larger than $1$ has significant practical merits. In our tests, we often used $\gamma=2$ as it yields an attractive time/space trade-off during construction and queries. 


\subsection{Parallelization performance}
\label{ssec:parallelization}

We evaluated the capability of our implementation to make use of multiple CPU cores. In Figure~\ref{fig:gamma} (right), we report the construction times with respect to the number of threads. 
We observe a near-ideal speed-up with respect to the number of threads with diminishing returns when using more than 10 threads, which is likely due to cache misses that induce a memory access bottleneck.


In addition to these results, 
we applied \bb on a key set of 10 billion keys and on a key set of 100 billion keys, again using default parameters and 8 threads. The memory usage was respectively 4.96GB and 49.49GB, and the construction time was respectively 462 seconds and 8913 seconds, showing the scalability of \bb.

\subsection{Comparisons with state of the art methods}
\label{ssec:sota}
We compared \bb with state-of-the-art \hf methods. CHD (\url{http://cmph.sourceforge.net/}) is 
an implementations of the compressed
hash-and-displace algorithm~\cite{belazzougui2009hash}. EMPHF~\cite{belazzougui2014cache} is based on random hypergraph peeling, and the HEM~\cite{botelho2013practical} implementation in EMPHF is based on partitioning the input data; both methods use external memory during construction.  
We did not perform comparisons with similar techniques as ours~\cite{chapman2011meraculous,lu2006perfect,Muller2014}, given that stand-alone implementations were not available. 
Our benchmark code is available at \url{https://github.com/rchikhi/benchmphf}. 

\begin{figure}[h]
  \centering
  \begin{tabular}{c}
    \includegraphics[width=0.75\textwidth]{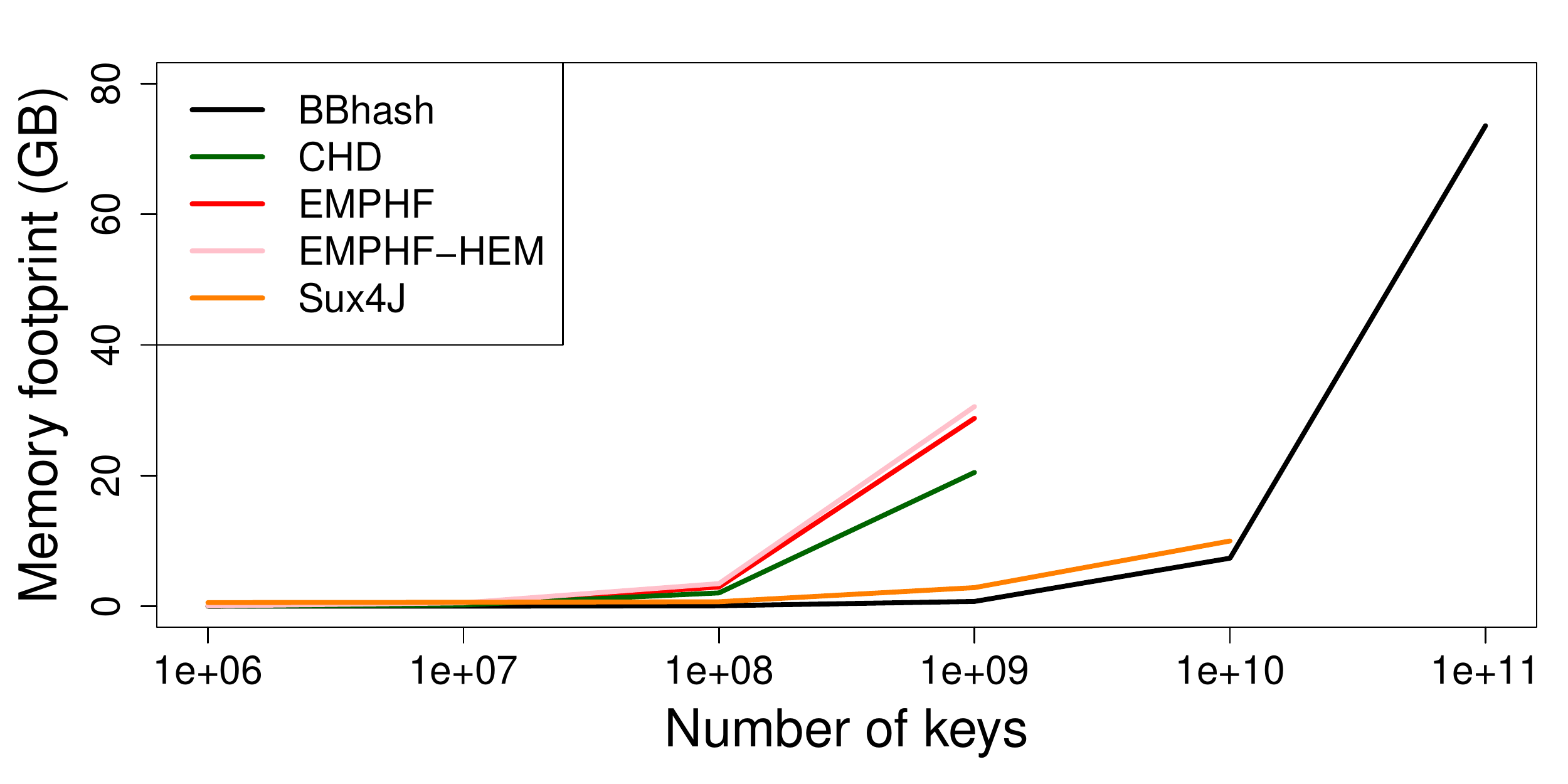}\\
  \includegraphics[width=0.75\textwidth]{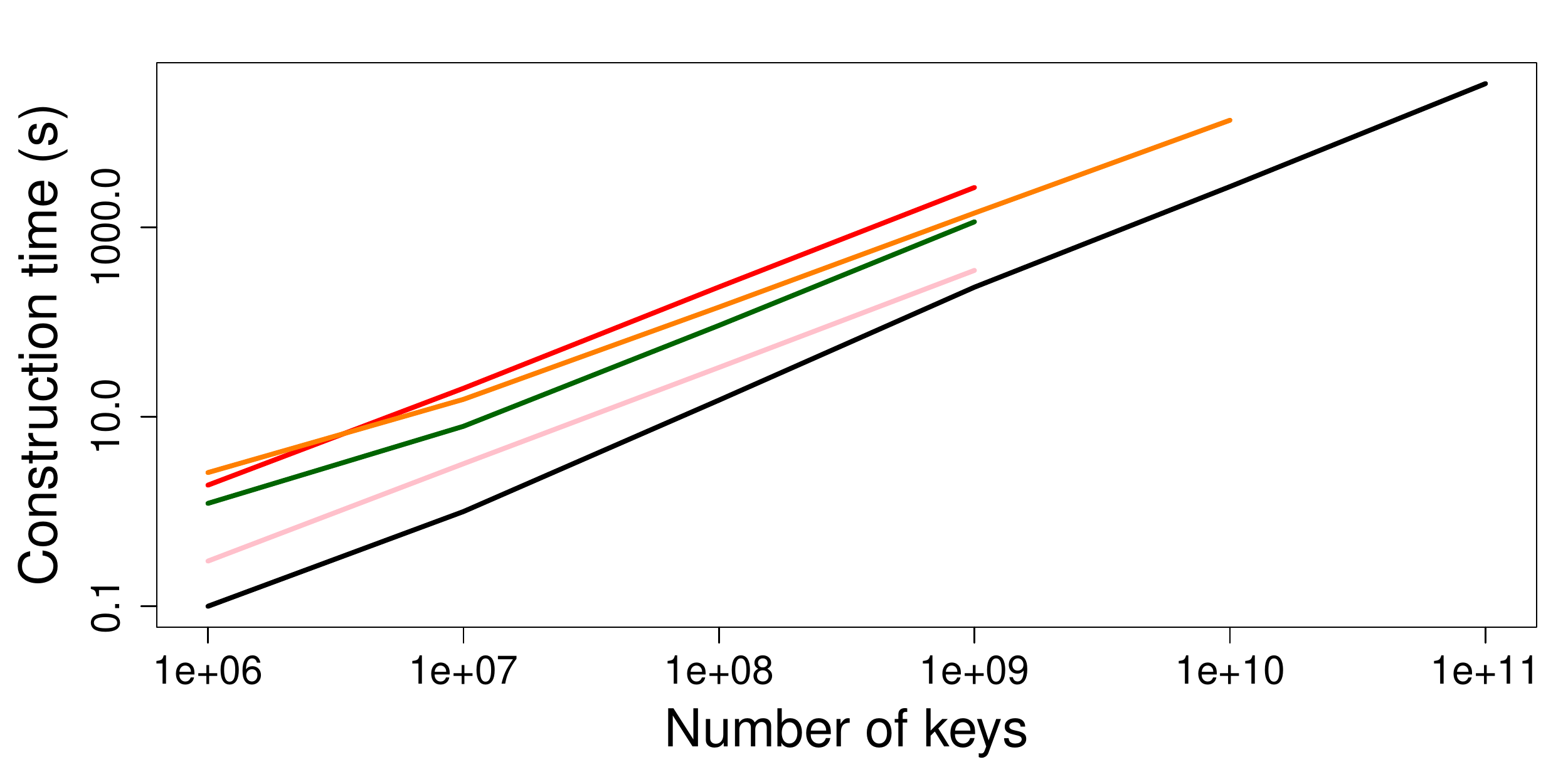}   
  \end{tabular}

  \caption{Memory footprint and construction time with respect to the number of keys. 
  All libraries were run using default parameters, including $\gamma=2$ for \bb. For a fair comparison, \bb was executed on a single CPU thread.  Except for Sux4J, missing data points correspond to runs that exceeded the amount of available RAM. Sux4J limit comes from the disk usage, estimated at approximately 4TB for $10^{11}$ keys. }
  \label{fig:scala}
\end{figure}  


\begin{table}[h]
\centering
\begin{tabular}{l|c|c|c|c|c}
Method &\begin{tabular}{c} Query \\time (ns)\end{tabular} & \begin{tabular}{c}\hf size\\ (bits/key)\end{tabular} & \begin{tabular}{c}  Const.\\time$^*$\\(s)\\ \end{tabular} &  
\begin{tabular}{c} Const.\\  memory$^{**}$\\  \end{tabular} & \begin{tabular}{c}Disk.\\usage\\(GB)\end{tabular}\\ 
\hline
\bb  $\gamma=1$  & 271 &  3.1 & 60 (393) & 3.2 (376) & 8.23\\ 
\bb  $\gamma=1$ minirank & 279 &  2.9 & 61(401) &  3.2 (376) &8.23\\
\bb  $\gamma=2$   & 216 &  3.7 & 35 (229) & 4.3 (516) &4.45\\ 
\bb  $\gamma=2$ nodisk   & 216 &  3.7 & 80 (549) &  6.2 (743) &0\\
\bb  $\gamma=5$   & 179 &  6.9 & 25 (162) & 10.7 (1,276) &1.52\\  
EMPHF  & 246 &   2.9 & 2,642 & 247.1 (29,461)$\dagger$  & 20.8\\
EMPHF HEM  & 581 &  3.5 & 489 & 258.4 (30,798)$\dagger$ & 22.5 \\
CHD  & 1037 &  2.6 & 1,146 & 176.0 (20,982) & 0\\
Sux4J & 252 & 3.3 & 1,418 & 18.10 (2,158) & 40.1	
\end{tabular} 
\caption{Performance of different \hf algorithms applied on a key set composed of $10^9$ 64-bits random integers, of size 8GB.   Each time result is the average value over three tests. The 'nodisk' row implements the second strategy described in Section~\ref{ssec:diskusage}, and the 'minirank' row samples ranks every 1024 positions instead of 512 by default. $^*$The column ``\textit{Const. time}'' indicates the construction time in seconds. In the case of \bb, the first value is the construction time using eight CPU threads and the second value in parenthesis is the one using one CPU thread. $^{**}$The column ``\textit{Const. memory}'' indicates the RAM used during the \hf construction, in bits/key and the total in MB in parenthesis. $\dagger$ The memory usages of EMPHF and EMPHF HEM reflect the use of memory-mapped files (\texttt{mmap} scheme).}
\label{MPHFPERF}
\end{table}



Figure~\ref{fig:scala} shows that all evaluated methods are able to construct {\hf}s that contain a billion of elements, while only \bb scales up to datasets that contain $10^{11}$ elements and more. Overall, \bb shows consistently better time and memory usage during construction. 

We additionally compared the resulting \hf size, i.e. the space of the data structure returned by the construction algorithm, and the mean query time across all libraries on a dataset consisting of a billion keys (Table~\ref{MPHFPERF}). 
{\hf}s produced by \bb range from 2.89 bits/key (when $\gamma=1$ and ranks are sampled every 1024 positions) to 6.9 bits/key (when $\gamma=5$ and a rank sampling of 512). The 0-0.8 bits/key size difference between our implementation and the theoretical space of \bb structure size is due to additional space used by the rank structure. We believe that a reasonable compromise in terms of query time and structure size is 3.7 bits/key with $\gamma=2$ and a rank sampling of 512, which is marginally larger than the \hf sizes of other libraries (ranging from 2.6 to 3.5 bits/key). As we argued in the Introduction that using 1 more bit per key is an acceptable trade-off for performance.

Construction times vary by one or two orders of magnitude across methods, \bb being the fastest. 
With default parameters ($\gamma=2$, rank sampling of 512), \bb has a construction memory footprint 40$\times$ to 60$\times$ smaller than other libraries except for Sux4j, for which \bb remains 4$\times$ smaller. 
Query times 
are roughly within an order of magnitude $(179-1037 \text{ ns})$ of each other across methods, with a slight advantage for \bb when $\gamma \geq 2$.
Sux4j achieves an attractive balance with low construction memory and query times, but high disk usage. In our tests, the high disk usage of Sux4j was a limiting factor for the construction of very large {\hf}s.

Note that EMPHF, EMPHF HEM and Sux4j implement a disk partitioning strategy, that could in principle also be applied to others methods, including ours. Instead of creating a single large \hf, they partition the set of input keys on disk and construct many small {\hf}s independently. In theory this technique allows to engineer the \hf construction algorithm to use parallelism and lower memory, at the expense of higher disk usage. In practice we observe that the existing implementations that use this technique are not parallelized. While EMPHF en EMPHF HEM used relatively high memory in our tests (around 30 GB for 1 billion elements) due to memory-mapped files, they also completed the construction successfully on another machine that had 16 GB of available memory. However, we observed what appears to be limitations in the scalability of the scheme: we were unable to run EMPHF and EMPHF HEM on an input of 100 billion elements. Regardless, we view this partitioning technique as promising but orthogonal to the design of efficient "monolithic" {\hf}s constructions such as \bb.

\subsection{Performance on an actual dataset}
\label{ssec:realdata}
In order to ensure that using pseudo-random integers as keys does not bias results, we ran \bb using string as keys. We used n-grams extracted from the Google Books Ngram dataset\footnote{http://storage.googleapis.com/books/ngrams/books/datasetsv2.html}, version 20120701. In average the n-gram size is 18. We also generated random words of size 18. As reported in Table~\ref{real}, we obtained highly similar results than those obtained with random integer keys. 

\begin{table}[h]
\centering
\setlength\tabcolsep{1pt}
\begin{tabular}{c|c|c|c}
Dataset &\begin{tabular}{c} Query time (ns)\end{tabular} & \begin{tabular}{c}\hf size\\ (bits/key)\end{tabular} & \begin{tabular}{c}  Const. time\\(s) \end{tabular}  \\
\hline
$10^8$ Random strings & 325 & 3.7 & 35\\
$10^8$ Ngrams & 296 &   3.7 & 37\\
\end{tabular} 
\caption{Performance of \bb ($\gamma=2$, 8 threads) when using ASCII strings as keys.}
\label{real}
\end{table}

\subsection{Indexing a trillion keys}
\label{ssec:huge}
We performed a very large-scale test by creating an \hf for $10^{12}$ keys. For this experiment, we used a machine with 750 GB of RAM. Since storing that many keys would require 8 TB of disk space, we instead used a procedure that generates deterministically a stream of $10^{12}$ pseudo-random integers in $[0,2^{64}-1]$. We considered the streamed values as input keys without writing them to disk. Thus, the reported computation time should not be compared to previously presented results as this experiment has no disk accesses.
The test was performed using $\gamma=2$, 24 threads, and keys were loaded in memory when $|F_i|\leq 2\%$ of total keys (i.e. when remaining number of keys to index was lower than 20 billion).

Creating the \hf took $35.4$ hours and required $637$ GB RAM. This memory footprint is roughly separated between the bit arrays ($\approx 459$ GB) and the memory required for loading 20 billion keys in memory ($\approx 178$ GB).
The final \hf occupied $3.71$ bits per key.


\section{Conclusion}
We propose a resource-efficient and highly scalable algorithm for constructing and querying {\hf}s.
Our algorithmic choices were motivated by simplicity: the method only relies on bit arrays and classical hash functions. 
While the idea of recording collisions in bit arrays to create {\hf}s is not novel~\cite{chapman2011meraculous,lu2006perfect}, to the best of our knowledge \bb is the first implementation that is competitive with the state of the art.
The construction is particularly time-efficient as it is parallelized and mainly consists in hashing keys and performing memory accesses. Moreover, the additional data structures used during construction are provably small enough to ensure a low memory overhead during construction. In other words, creating the \hf does not require much more space than the resulting \hf itself. This aspect is important when constructing {\hf}s on large key sets in practice.

Experimental results show that \bb generates {\hf}s that are slightly larger than those produced by other methods. However \bb is by far the most efficient in terms of construction time, query time, memory and disk footprint for indexing large key sets (of cardinality above $10^9$ keys). The scalability of our approach was confirmed by constructing {\hf}s for sets as large as $10^{12}$ keys. To the best of our knowledge, no other \hf implementation has been tested on that many keys. 

A time/space trade-off is achieved through the $\gamma$ parameter. The value $\gamma=1$ yields {\hf}s that occupy roughly $3N$ bits of space and have little memory overhead during construction. Higher $\gamma$ values use more space for the construction and the final structure size, but they achieve faster construction and query times. 
Our 
results suggest that $\gamma=2$ is a good time-versus-space compromise, using 3.7 bits per key. 
With respect to hypergraph-based methods~\cite{belazzougui2014cache,botelho2007simple,botelho2013practical,Genuzio2016}, \bb offers significantly better construction performance, but the resulting \hf size is up to 1 bit/key larger. We however argue that the \hf size, as long as it is limited to a few bits per key, is generally not a bottleneck as many applications use {\hf}s to associate much larger values to keys. 
Thus, we believe that this work will unlock many HPC applications where the possibility to index billions keys and more is a huge step forward.

An interesting future work is to obtain more space-efficient {\hf}s using our method. We believe that a way to achieve this goal is to slightly change the hashing scheme. We would like to explore an idea inspired by the CHD algorithm for testing several hash functions at each level and selecting (then storing) one that minimizes the number of collisions. At the price of longer construction times, we anticipate that this approach could significantly decrease the final structure size. 

\section*{Acknowledgments}
This work was funded by French ANR-12-BS02-0008 Colib'read project. 
We thank the GenOuest BioInformatics Platform that provided the computing resources necessary for benchmarking. We thank Djamal Belazzougui for helpful discussions and pointers.

\bibliographystyle{plain}
\bibliography{references}
\newpage
\section*{Appendix}

\subsection*{Proofs of \hf size and memory required for construction}

\begin{proof}[MPHF size with $\gamma=1$]
\begin{align*}
\sum_{d\geq 0} |A_d| &= N \sum_{d\geq 0} (1-e^{-1})^{d}\\
&= N\frac{1}{1-(1-e^{-1})} &\text{   as } \lim_{d \to +\infty} (1-e^{-1})^d=0\\
&= eN
\end{align*}
\end{proof}

\begin{proof}[MPHF size using any $\gamma\geq1$]

$\text{With }\gamma\geq1:  |A_{d}| = \gamma|A_{d-1}| (1-e^{\frac{-1}{\gamma}}) =  \gamma|A_0| (1-e^{\frac{-1}{\gamma}})^{d} = \gamma N(1-e^{\frac{-1}{\gamma}})^{d}$\\
$$\text{Thus,  }\sum_{d\geq 0} |A_d| = \gamma N \sum_{d\geq 0} (1-e^{\frac{-1}{\gamma}})^{d}$$
Moreover, as $\lim_{d \to +\infty} (1-e^{\frac{-1}{\gamma}})^d=0 \text{ since for } \gamma>0, 0<1-e^{\frac{-1}{\gamma}}<1$, on has: 
$$\sum_{d\geq 0} |A_d| = \gamma N\frac{1}{1-(1-e^{\frac{-1}{\gamma}})} = \gamma e^{\frac{1}{\gamma}}N $$
\end{proof}

Note that this proof stands for any $\gamma$ value $>0$, but that with $\gamma<1$ the theoretical and practical \hf sizes increase exponentially as $\gamma$ get close to zero.

\begin{proof}[Lemma 1]
Let $m(d)$ be memory required during level $d$ and let $R$ be the ratio between the maximal memory needed during the \hf construction and the \hf total size denoted by $S$. Formally, $$R=\frac{\max_{d\geq0}(m(d))}{S}=\frac{\max_{d\geq0}(m(d))}{\gamma e^{\frac{1}{\gamma}}N}$$


\noindent
First we prove that $\lim_{d\rightarrow \infty} \frac{m(d)}{S} = 1$. 


\begin{align*}
m(d)=\sum_{i<d}|A_i|+2|A_d| = \gamma N\left(\frac{1-(1-e^{\frac{-1}{\gamma}})^d}{e^{\frac{-1}{\gamma}}} + 2(1-e^{\frac{-1}{\gamma}})^d\right)\\
\end{align*}
Since for $\gamma>0$, $0<1-e^{\frac{-1}{\gamma}}<1$, then
$
\lim_{d\rightarrow \infty} m(d) = \gamma e^{\frac{1}{\gamma}}N
$. Thus $\lim_{d\rightarrow \infty} \frac{m(d)}{S} = 1$.

\noindent
Before going further, we need to compute $m(d+1)-m(d)$: 
\begin{align*}
m(d+1)-m(d)&=\sum_{i<d+1}|A_i|+2|A_{d+1}|-\sum_{i<d}|A_i|+2|A_d|\\
           &=|A_d|+2|A_{d+1}|-2|A_d|= 2|A_{d+1}|-|A_d|\\
           &=2\gamma N(1-e^{\frac{-1}{\gamma}})^{d+1} - \gamma N(1-e^{\frac{-1}{\gamma}})^{d}\\
           &=\gamma N(1-e^{\frac{-1}{\gamma}})^{d} (2(1-e^{\frac{-1}{\gamma}})-1)\\
           &=\gamma N(1-e^{\frac{-1}{\gamma}})^{d} (1-2e^{\frac{-1}{\gamma}})
\end{align*}
We now prove $R\leq1$ when $\gamma\leq\frac{1}{\log(2)}$ and also, $R<2$ when $\gamma>\frac{1}{\log(2)}$. 
\begin{itemize}
\item Case 1: $\gamma\leq\frac{1}{\log(2)}$

We have $ \frac{m(0)}{S} = 2e^{-\frac{1}{\gamma}} \leq 2e^{-\log(2)} = 1$.

Moreover, as $m(d+1)-m(d)=\gamma N(1-e^{\frac{-1}{\gamma}})^{d} (1-2e^{\frac{-1}{\gamma}})$ and as, with  $\gamma\leq\frac{1}{\log(2)}$: $1-e^{\frac{-1}{\gamma}} \geq 0.5$, and $1-2e^{\frac{-1}{\gamma}} \geq 0$ then $m(d+1)-m(d)\geq 0$, thus, $m$ is an increasing function. 

To sum up, with $\gamma\leq\frac{1}{\log(2)}$, we have \textbf{1/} that $\frac{m(0)}{S} \leq 1$,  \textbf{2/} that $\lim_{d\rightarrow \infty} \frac{m(d)}{S} = 1$, and \textbf{3/} that $m$ is increasing, then $R\leq1$.

\item Case 2: $\gamma>\frac{1}{\log(2)}$
We have $ \frac{m(0)}{S} = 2e^{-\frac{1}{\gamma}}$. With $\gamma>\frac{1}{\log(2)}$,  $1<\frac{m(0)}{S}<2$.  
Moreover, $m(d+1)-m(d)=\gamma N(1-e^{\frac{-1}{\gamma}})^{d} (1-2e^{\frac{-1}{\gamma}})$ is negative as: $1-e^{\frac{-1}{\gamma}}>0$ and $1-2e^{\frac{-1}{\gamma}}<0$ for $\gamma>\frac{1}{\log(2)}$.
Thus $m$ is a decreasing function with $d$. 

With $\gamma>\frac{1}{\log(2)}$, we have \textbf{1/} that $\frac{m(0)}{S}<2$, \textbf{/2} that $\lim_{d\rightarrow \infty} \frac{m(d)}{S} = 1$ and  \textbf{/3} that $m$ is decreasing. Thus $R <2$.
\end{itemize}
\end{proof}
\vfill

\subsection*{Algorithms pseudo-codes}

\begin{algorithm}[H]
 \KwData{$F_0$ a set of $N$ keys, integers $\gamma$ and $last
$}
 \KwResult{array of bit arrays $\{A_0,A_1,\ldots, A_{last}\}$, hash table $H$}
 i=0\;
 \While{$F_i$ not empty and $i\leq last$}{
    $A_i=ArrayFill(F_i,\gamma)$\;
    \ForEach{key $x$ of $F_i$}{
        $h=hash(x)\mod(\gamma*N)$\;
        \If{$A_i[h]==0$}{$F_{i+1}.add(x)$}
    }
    i=i+1\;
 }
 Construct $H$ using remaining elements from $F_{last+1}$\;
 Return $\{A_0,A_1,\ldots,A_{last},H\}$
 \caption{\hf construction.}
 \algorithmfootnote{In practice $F_i$ with $i>1$ are stored on disk (see Section~\ref{ssec:diskusage}). The hash table $H$ ensures that elements in $F_{last+1}$ are mapped without collisions to integers in $[|F_0|-|F_{last+1}|+1,|F_0|]$}
 \label{construction}
\end{algorithm}

\begin{algorithm}[H]
 \KwData{$F$ array of $N$ keys, integer $\gamma$}
 \KwResult{ bit array  $A$}
 Zero-initialize $A$ and $C$ two bit arrays with $\gamma*N$ elements\;
 \ForEach{key $x$ of $F$}{
    $h=hash(x) \mod (\gamma*N)$\;
    \If{$A[h]==0$ and $C[h]==0$}{$A[h]=1$\;}
    \If{$A[h]==1$ and $C[h]==0$}{$A[h]=0$\; $C[h]=1$\;}
    \If{$A[h]==0$ and $C[h]==1$}{Skip\;}
 }
 Delete $C$\;
 Return $A$\;
 \caption{$ArrayFill$}
 \algorithmfootnote{Note that the case $A[h]==1$ and $C[h]==1$ never happens.}
 \label{AlgoFill}
\end{algorithm}

\begin{algorithm}[H]
 \KwData{bit arrays $\{A_0,A_1,\ldots, A_{\text{last}}\}$, hash table $H$, key $x$}
 \KwResult{integer index of $x$}
  i=0\;
  \While{$i \leq last$}{
    $h=hash_i(x) \mod A_i.size()$\;
    \If{$A_i[h]==1$}{
    \Return{$\sum_{j<i}{|A_j|} +rank(A_i[h])$} \;
    }
    $i=i+1$\;
    }
     \Return{H[x]} \;
 \caption{\hf query}
 \algorithmfootnote{Note, when $x$ is not an element from the key set of the \hf, the algorithm may return a wrong integer index.}
  \label{query}
\end{algorithm}

\section*{Commands}
In this section we describe used commands for each presented result.
Time and memory usages where computed using ``\textit{/usr/bin/time --verbatim}'' unix command. The disk usage was computed thanks to a home made script measuring each 1/10 second the size of the directory using the ``\textit{du -sk''} unix command, and recording the highest value. The \bb library and its \textit{Bootest} tool are available from \url{https://github.com/rizkg/BBHash}.

\paragraph*{Commands used for Section~\ref{ssec:gamma}:} 
\begin{verbatim}
for ((gamma=1;gamma<11;gamma++)); do 
	./Bootest 1000000000 1 ${gamma} -bench  
done
\end{verbatim}
Note that 1000000000 is the number of keys tested and 1 is the number of used cores.

Additional tests, with larger key set and 8 threads: 
\begin{verbatim}
for ((gamma=1;gamma<11;gamma++)); do 
	./Bootest 1000000000 1 ${gamma} -bench  
done
\end{verbatim}

\paragraph*{Commands used for Section~\ref{ssec:parallelization}:}

\begin{verbatim}
for keys in 10000000000 100000000000; do 
	./Bootest ${keys} 8 2 -bench  
done
\end{verbatim}

\paragraph*{Commands used for Section~\ref{ssec:sota}:}
We remind that our benchmark code, testing EMPHF, EMPHF MEM, CHD, and Sux4J is available at \url{https://github.com/rchikhi/benchmphf}. 
\begin{itemize}
\item \bb commands:
\begin{verbatim}
for keys in 1000000 10000000 100000000 10000000000\
 10000000000 100000000000; do 
	./Bootest ${keys} 1 2 -bench  
done
\end{verbatim}
\item \bb command with nodisk (Table~\ref{MPHFPERF}) was 
\begin{verbatim}
./Bootest 1000000000 1 2 -bench -nodisk
\end{verbatim}
and
\begin{verbatim}
./Bootest 1000000000 8 2 -bench -nodisk
\end{verbatim}
respectively for one and height threads.
Other commands from Table~\ref{MPHFPERF} were deduced from previously presented \bb computations. 

\item Commands EMPHF \& EMPHF HEM: 
\begin{verbatim}
for keys in 1000000 10000000 100000000 10000000000\
 10000000000 100000000000; do 
	./benchmphf ${keys} -emphf
done
\end{verbatim}
EMPHF (resp. EMPHF HEM) is tested by using the \#define EMPHF\_SCAN macro (resp. \#define EMPHF\_HEM).  In order to assess the disk size footprint, the line ``\textit{unlink(tmpl);}'' from file ``\textit{emphf/mmap\_memory\_model.hpp}'' was commented.

\item Commands CHD:
\begin{verbatim}
for keys in 1000000 10000000 100000000 10000000000\
 10000000000 100000000000; do 
	./benchmphf ${keys} -chd
done
\end{verbatim}

\item Commands Sux4J: 

for each size, the \textit{``Sux4J/slow/it/unimi/dsi/sux4j/mph/LargeLongCollection.java''} was modified indicating the used size. 
\begin{verbatim}
./run-sux4j-mphf.sh
\end{verbatim}

\paragraph*{Commands used for Section~\ref{ssec:realdata}:}
As explained Section~\ref{ssec:realdata}, the keyString.txt file is composed of n-grams extracted from the Google Books Ngram dataset\footnote{http://storage.googleapis.com/books/ngrams/books/datasetsv2.html}, version 20120701.
\begin{verbatim}
./BootestFile keyStrings.txt 10 2
\end{verbatim}

\paragraph*{Commands used for Section~\ref{ssec:huge}:}
 \bb command for indexing a trillion keys, with keys generated on the fly.
\begin{verbatim}
./Bootest 1000000000000 24 2 -onthefly
\end{verbatim}

\end{itemize}
\end{document}